\documentclass[11pt]{article}
\usepackage{amsmath, amssymb, amsfonts, amsthm}

\newtheorem{theorem}{Theorem}[section]

\newtheorem{corollary}[theorem]{Corollary}

\newcommand{\rd}{{\rm d}}
\newcommand{\be}{\begin{equation}}
\newcommand{\ee}{\end{equation}}
\newcommand{\bey}{\begin{eqnarray}}
\newcommand{\eey}{\end{eqnarray}}

\newcommand{\eps}{\varepsilon}

\newcommand{\ph}{\varphi}

\newcommand{\cU}{{\cal U}}

\newcommand{\bR}{{\mathbb R}}
\newcommand{\bC}{{\mathbb C}}
\newcommand{\bN}{{\mathbb N}}

\newcommand{\tr}{\mbox{Tr}}

\newcommand{\cF}{{\cal F}}

\newcommand{\cH}{{\cal H}}
\newcommand{\cL}{{\cal L}}
\newcommand{\cN}{{\cal N}}

\input epsf

\newcommand{\donothing}[1]{}

\begin{document}

\title{Effective Evolution Equations in Quantum Physics}

\author{Benjamin Schlein\footnote{Partially supported by an ERC Starting Grant} \\ \\ Institute for Applied Mathematics, University of Bonn \\ Endenicher Allee 60, 53115 Bonn}

\maketitle

\begin{abstract}
In these notes, we review some recent mathematical results concerning the derivation of effective evolution equations from many body quantum mechanics. In particular, we discuss the emergence of the Hartree equation in the so-called mean field regime (for example, for systems of gravitating bosons), and we show that the Gross-Pitaevskii equation approximates the dynamics of initially trapped Bose-Einstein condensates. We explain how effective evolution equations can be derived, on the one hand, by analyzing the so called BBGKY hierarchy, describing the time-evolution of reduced density matrices, and, on the other hand, by studying the dynamics of coherent initial states in a Fock-space representation of the many body system.
\end{abstract}

\section{Introduction}
\label{sec:intro}
\setcounter{equation}{0}

Systems of interest in physics are typically characterized by a very large number of interacting degrees of freedom. Direct application of fundamental theories to study the properties of these systems is impossible. One of the main goals of
statistical mechanics consists therefore in the derivation of simpler effective theories, providing a good approximation to the fundamental equations in the relevant regimes.

We consider here a quantum mechanical system of $N$ particles moving in the three dimensional space and described by a wave function $\psi_N \in L^2 (\bR^{3N}, dx_1 \dots dx_N)$. Physically, the absolute value squared $|\psi_N (x_1, \dots, x_N)|^2$ determines the probability density for finding particle one at $x_1$, particle two at $x_2$ and so on. More generally, an observable of the system is a self-adjoint operator $A$ acting on the  $L^2 (\bR^{3N}, dx_1 \dots dx_N)$; the physical expectation of this observable in the state described by the wave function $\psi_N$ is given by the $L^2$-inner product $\langle \psi_N , A \psi_N \rangle$. The positions of the $N$ particles are associated with multiplication operators (the operator $A = x_j$, acting by multiplication with $x_j$, measures the position of the $j$-th particle). The momenta of the particles are associated with differential operators (the operator $A = -i \nabla_{x_j}$ measures the momentum of the $j$-th particle).

In nature, there exist two different types of particles, known as bosons and fermions, which behave differently with respect to permutations of the particles. The wave functions of bosonic systems are symmetric with respect to any permutation of the $N$ particles (in the sense that $\psi_N (x_{\pi 1}, \dots , x_{\pi N}) = \psi_N (x_1, \dots , x_N)$ for any permutation $\pi \in S_N$). The wave functions of fermionic systems, on the other hand, are antisymmetric with respect to permutations ($\psi_N (x_{\pi 1}, \dots , x_{\pi N}) = \sigma_{\pi} \psi_N (x_1, \dots , x_N)$, where $\sigma_\pi$ is the sign of the permutation $\pi \in S_N$). In these notes, we will exclusively consider bosonic systems, and therefore we will always assume the wave function $\psi_N$ to be symmetric with respect to permutations. We will denote by $L^2_s (\bR^{3N})$ the subspace of $L^2 (\bR^{3N})$ consisting of all permutation symmetric functions.

The evolution of the quantum system is governed by the $N$-particle Schr\"odinger equation
\begin{equation}\label{eq:schr} i\partial_t \psi_{N,t} = H_N \psi_{N,t} \end{equation}
for the time dependent wave function $\psi_{N,t}$ (the subscript $t$ just indicates the time dependence of the wave function; time derivatives will always be written as $\partial_t$). On the r.h.s. of the Schr\"odinger equation, $H_N$ is a self-adjoint operator acting on $L^2_s (\bR^{3N})$ and commonly known as the Hamilton operator (or Hamiltonian) of the system. It typically has the form
\begin{equation}\label{eq:ham} H = \sum_{j=1}^N \left(-\Delta_{x_j} + V_{\text{ext}} (x_j) \right) + \lambda \sum_{i<j}^N V (x_i -x_j) \end{equation}
where $V_{\text{ext}}$ is an external potential, $\lambda \in \bR$ is a coupling constant and $V(x_i -x_j)$ describes the interaction between particles $i$ and $j$. Here, and in the following, we restrict our attention to Hamilton operators with two-body potentials, neglecting interactions which depend at the same time on three or more particles.

The Schr\"odinger equation (\ref{eq:schr}) is a linear equation and its unique solution can be obtained by applying the unitary group generated by the self-adjoint operator $H_N$ to the initial wave function $\psi_{N,t=0}$, i.e. $\psi_{N,t} = e^{-iH t} \psi_{N,0}$. in other words, the well-posedness of the $N$-particle Schr\"odinger equation (\ref{eq:schr}) is not an issue (although, depending on the properties of the potentials $V_{\text{ext}}$, $V$, it may not be easy to show that the unbounded operator $H$, defined on an appropriate domain, is self-adjoint). When studying systems of interest in physics, however, it is generally very difficult to extract qualitative or quantitative information (beyond its well-posedness) from (\ref{eq:schr}). The reason is that, in typical situations, the number of particles involved in the dynamics is huge. It varies between $N \simeq 10^3$ in extremely dilute samples of Bose-Einstein condensates, to $N \simeq 10^{23}$ in chemical samples, up to $N \gtrsim 10^{45}$ in stars and other astronomical and cosmological systems. For such values of $N$, it is impossible to solve (\ref{eq:schr}) directly. Instead, it is desirable to derive effective equations approximating the dynamics governed by (\ref{eq:schr}) in the relevant interesting regimes.

Probably the simplest non-trivial situation where it is possible to obtain an effective approximation of the dynamics (\ref{eq:schr}) is the so-called mean field limit.
The mean field regime is characterized by a large number of particles which interact very weakly with all other particles in the system, so that, effectively, the two-body interaction experienced by each particle can be approximated by an averaged, mean field, potential. The Hamiltonian (\ref{eq:ham}) describes a mean field regime if $N \gg 1$ (large number of particles) and $\lambda \ll 1$ (weak coupling), so that $N\lambda =: \kappa$ remains of order one (this guarantees that the total potential, the total force acting on each particle is of order one). To study the evolution in the mean field regime we consider a factorized initial wave function $\psi_N = \ph^{\otimes N}$ (meaning that $\psi_N (x_1, \dots , x_N) = \prod_{j=1}^N \ph (x_j)$) for an arbitrary one-particle orbital $\ph \in L^2 (\bR^3)$ and we study its evolution $\psi_{N,t} = e^{-i H_N t} \psi_{N,0}$, as generated by the mean field Hamilton operator
\begin{equation}\label{eq:mf-ham} H_N = \sum_{j=1}^N \left( -\Delta_{x_j} + V_{\text{ext}} (x_j) \right) + \frac{\kappa}{N} \sum_{i<j} V (x_i - x_j) \end{equation}
in the limit of large $N$. Because of the interaction, factorization cannot be preserved by the time-evolution. Nevertheless, because of the mean field character of the Hamiltonian, we may expect that factorization is approximately preserved in the limit of large $N$. If this is true, we would have
\begin{equation}\label{eq:conv} \psi_{N,t} (x_1, \dots , x_N) \simeq \prod_{j=1}^N \ph_t (x_j) \end{equation}
in an appropriate sense and for an appropriately evolved one-particle orbital $\ph_t \in L^2 (\bR^3)$. Under this assumption, it is simple to find a self-consistent equation for the evolution of the orbital $\ph_t$. In fact, if (\ref{eq:conv}) holds true, the particles are approximately distributed in space independently of each others, with the probability density $| \ph_t|^2$. The total potential experienced by, say, particle $j$, can therefore be approximated by
\[ \frac{\kappa}{N} \sum_{i\not = j} V (x_i -x_j) \simeq \frac{\kappa}{N} \sum_{i\not = j} \int dy \, |\ph_t (y)|^2 V (y -x_j) \simeq \kappa (V * |\ph_t|^2) (x_j) \, . \]
We conclude that the one-particle orbital $\ph_t$ must evolve according to the self-consistent, non-linear, Hartree equation
\begin{equation}\label{eq:hartree} i\partial_t \ph_t = -\Delta \ph_t + V_{\text{ext}} \ph_t + \kappa (V*|\ph_t|^2) \ph_t \, . \end{equation}

To understand in which sense (\ref{eq:conv}) can be correct, we introduce the concept of reduced density matrices associated with a bosonic wave function $\psi_{N,t}$. We define, first of all, the density matrix associated with $\psi_{N,t}$ as the orthogonal projection $\gamma_{N,t} = |\psi_{N,t} \rangle \langle \psi_{N,t}|$ onto $\psi_{N,t}$ (we use here the shorthand notation $|\psi \rangle \langle \psi|$ to denote the orthogonal projection onto $\psi$). For $k=1,\dots , N$, we define then the $k$-particle reduced density matrix $\gamma^{(k)}_{N,t}$ by taking the partial trace of $\gamma_{N,t}$ over the last $N-k$ variables, i.e.
\begin{equation}\label{eq:redk-1} \gamma^{(k)}_{N,t} = \tr_{k+1, \dots , N} \, \gamma_{N,t} \, . \end{equation}
In other words, the $k$-particle reduced density matrix $\gamma^{(k)}_{N,k}$ is defined as the non-negative, trace-class operator on $L^2 (\bR^{3k})$ with the integral kernel
\begin{equation}\label{eq:redk-2}\begin{split}
\gamma^{(k)}_{N,t} & ( x_1, \dots , x_k ; x'_1,
\dots x'_k) \\ &= \int dx_{k+1} \dots dx_N \, \psi_{N,t} (x_1, \dots ,x_k, x_{k+1}, \dots , x_N) \overline{\psi}_{N,t} (x'_1, \dots , x'_k, x_{k+1} , \dots , x_N) \, . \end{split} \end{equation}
By definition $\tr \, \gamma^{(k)}_{N,t} = 1$ for all $k=1, \dots , N$. Using the $k$-particle reduced density $\gamma^{(k)}_{N,t}$ one can compute the expectation of $k$-particle observables (observables depending non-trivially on at most $k$ particles). In fact,
\[ \langle \psi_{N,t} , \left( O^{(k)} \otimes 1^{(N-k)} \right) \psi_{N,t} \rangle = \tr \, \left( O^{(k)} \otimes 1^{(N-k)} \right) \, \gamma_{N,t} = \tr \, O^{(k)}  \gamma^{(k)}_{N,t} \, . \]

It turns out that reduced density matrices provide the correct language to understand the convergence (\ref{eq:conv}) in the limit of large $N$.

\begin{theorem}\label{thm:mf}
Under appropriate assumptions on the potentials $V, V_{\text{ext}}$ (see discussion below), consider the factorized $N$-particle wave function $\psi_N = \ph^{\otimes N}$, for an arbitrary $\ph \in H^1 (\bR^3)$. Let $\psi_{N,t} = e^{-i H_N t} \psi_N$ be the evolution of $\psi_N$, as generated by the mean field Hamiltonian (\ref{eq:mf-ham}). Then, for any fixed $k \in \bN$ and $t \in \bR$, we have, as $N \to \infty$,
\begin{equation}\label{eq:conv2} \gamma^{(k)}_{N,t} \to |\ph_t \rangle \langle \ph_t |^{\otimes k} \end{equation}
where $\ph_t$ is the solution of the nonlinear Hartree equation (\ref{eq:hartree}) with initial data $\ph_{t=0} = \ph$. The convergence in (\ref{eq:conv2}) is in the trace class topology.
\end{theorem}

In particular, Theorem \ref{thm:mf} implies that, for any bounded $k$-particle observable $O^{(k)}$, and for any fixed $t\in \bR$, we have
\[ \left\langle \psi_{N,t} , \left( O^{(k)} \otimes 1^{(N-k)} \right) \psi_{N,t} \right\rangle \to \left\langle \ph_t^{\otimes k} , O^{(k)} \ph_t^{\otimes k} \right\rangle \] as $N \to \infty$. In other words, as long as we compute the expectation of observables depending only on a fixed number of particles, the solution of the full Schr\"odinger equation $\psi_{N,t}$ can be approximated, as $N \to \infty$, by the product $\ph_t^{\otimes N}$ of solutions of the Hartree equation (\ref{eq:hartree}).

The first proof of this result was obtained by Spohn in \cite{Sp}, for bounded interaction potential $V \in L^{\infty} (\bR^3)$. Later, Spohn's approach was extended by Erd\H os and Yau in \cite{EY} to interactions with Coulomb singularity $V(x) = \pm 1/|x|$ (partial results for the Coulomb case were also obtained by Bardos-Golse-Mauser in \cite{BGM}). Note that, in all these works, the important assumptions concern the interaction potential; only very minor conditions have to be imposed on the external potential $V_{\text{ext}}$ (to make sure that the operator $H_N$ is self-adjoint).

The main idea of the approach developed in \cite{Sp} and then extended in \cite{EY,BGM} is to study directly the time-evolution of the reduced density matrices. {F}rom the Schr\"odinger equation (\ref{eq:schr}) for the wave-function $\psi_{N,t}$ it is simple to derive a hierarchy, commonly known as the BBGKY hierarchy (BBGKY stands for Bogoliubov, Born, Green, Kirkwood, Yvon), consisting of $N$ coupled equations describing the evolution of the reduced densities. For $k =1, \dots , N$, we find
\begin{equation}\label{eq:BBGKY} \begin{split} i\partial_t \gamma^{(k)}_{N,t} = \; &\sum_{j=1}^k \left[ -\Delta_{x_j} + V_{\text{ext}} (x_j) , \gamma^{(k)}_{N,t}\right] + \frac{1}{N} \sum_{i<j}^k \left[ V (x_i - x_j) , \gamma^{(k)}_{N,t} \right] \\ &+ \left( 1- \frac{k}{N} \right) \sum_{j=1}^k \tr_{k+1} \, \left[ V (x_j - x_{k+1}), \gamma^{(k+1)}_{N,t} \right] \, . \end{split} \end{equation}
Here, we use the convention that $\gamma^{(N+1)}_{N,t} = 0$. In the last term, $\tr_{k+1}$ indicates the partial trace over the $(k+1)$-th particle. It is simple to understand the origin of the three terms on the r.h.s. of (\ref{eq:BBGKY}). The first term corresponds to the kinetic energy of the first $k$ particles. The second term describes the interactions among the first $k$ particles. The last term, on the other hand, describes the interaction between the first $k$ particles and the other $(N-k)$ particles, which are integrated out in the definition of $\gamma^{(k)}_{N,t}$ (this is why the last term depends on $\gamma^{(k+1)}_{N,t}$ and not only on $\gamma^{(k)}_{N,t}$). As $N \to \infty$, the BBGKY hierarchy converges, at least formally, to an infinite hierarchy of equations. Suppose that $\{ \gamma_{\infty,t}^{(k)} \}_{k \geq 1}$ denotes a limit point of the sequence $\{ \gamma_{N,t}^{(k)} \}_{k =1}^N$ (with respect to an appropriate product topology). Then, from (\ref{eq:BBGKY}), we may expect the limit point to satisfy the infinite hierarchy ($k \geq 1$)
\begin{equation}\label{eq:inf-hier}  i\partial_t \gamma^{(k)}_{\infty,t} = \sum_{j=1}^k \left[ -\Delta_{x_j} + V_{\text{ext}} (x_j), \gamma^{(k)}_{\infty,t}\right] + \sum_{j=1}^k \tr_{k+1} \, \left[ V (x_j - x_{k+1}), \gamma^{(k+1)}_{\infty,t} \right] \,.\end{equation}
It is then worth noticing that this infinite hierarchy has a factorized solution. In fact, it is simple to check that the ansatz $\gamma^{(k)}_{\infty,t} = |\ph_t \rangle \langle \ph_t|^{\otimes k}$ solves (\ref{eq:inf-hier}) if and only if $\ph_t$ solves the nonlinear Hartree equation (\ref{eq:hartree}). This observation suggests that, in order to obtain a rigorous proof of Theorem \ref{thm:mf}, one can proceed as follows. First, one has to show the compactness of the sequence $\{ \gamma^{(k)}_{N,t} \}_{k = 1}^N$ with respect to an appropriate (weak) topology. In the second step, one has to identify the limit points $\{ \gamma^{(k)}_{\infty,t} \}_{k\geq 1}$ of the sequence $\{ \gamma^{(k)}_{N,t} \}_{k = 1}^N$ as solutions of the infinite hierarchy (\ref{eq:inf-hier}). Finally, one has to show the uniqueness of the solution of the infinite hierarchy (\ref{eq:inf-hier}). These three steps immediately imply the convergence of the sequence $\{ \gamma^{(k)}_{N,t} \}_{k =1}^N$ towards the factorized solution $\{ |\ph_t \rangle \langle \ph_t|^{\otimes k} \}_{ k \geq 1}$ with respect to the weak topology; since $|\ph_t \rangle \langle \ph_t|^{\otimes k}$ is a rank one projection, it is then easy to check that the weak convergence also implies strong convergence (in the trace-norm topology). Let us remark here that, as noticed by Chen and Pavlovi{\'c} in \cite{CP2}, this approach can also be extended to many-body systems interacting via three-body potentials (namely potentials depending at the same time on three particles).

\section{Dynamics of Bose-Einstein Condensates}
\label{sec:BEC}
\setcounter{equation}{0}

A trapped Bose gas can be described by the Hamilton operator
\begin{equation}\label{eq:trap-ham} H_N^{\text{trap}} = \sum_{j=1}^N \left( - \Delta_{x_j} + V_{\text{ext}} (x_j) \right) + \sum_{i<j}^N N^2 V (N (x_i - x_j)) \end{equation} acting on the Hilbert space $L^2_s (\bR^{3N}, dx_1, \dots  dx_N)$. Here $V_{\text{ext}}$ is a confining potential (with $V_{\text{ext}} (x) \to \infty$ as $|x| \to \infty$), and $V$ is a short range (possibly compactly supported), repulsive (meaning that $V(x) \geq 0$, for all $x\in \bR^3$), bounded interaction potential. In (\ref{eq:trap-ham}), the interaction potential scales with the number of particles $N$, so that its scattering length is of the order $1/N$. The scattering length is a physical quantity which measures the effective range of the potential $V$. It is defined through the solution of the zero-energy scattering equation
\begin{equation}\label{eq:0-en} \left( -\Delta + \frac{1}{2} V (x) \right) f = 0 \end{equation}
with the boundary condition $f (x) \to 1$ as $|x| \to \infty$. For large $|x|$, $f(x)$ has the form
\[ f (x) \simeq 1 - \frac{a_0}{|x|} + o (|x|^{-1}) \]
for an appropriate constant $a_0$, which is defined to be the scattering length of $V$. Equivalently, the scattering length $a_0$ can be defined through  \begin{equation}\label{eq:scatt} 8 \pi a_0 = \int dx \, V (x) f (x) \, . \end{equation} It is simple to check that, for repulsive potentials, $a_0 >0$. By scaling, it is also clear that, if $a_0$ denotes the scattering length of the potential $V$, the scattering length of the rescaled potential $V_N (x) = N^2 V (N x)$ appearing in (\ref{eq:trap-ham}) is given by $a = a_0 /N$.

In \cite{LSY}, Lieb, Seiringer and Yngvason proved that, in the limit of large $N$, the ground state energy of the Hamiltonian (\ref{eq:trap-ham}) can be approximated, for large $N$, minimizing the so called Gross-Pitaevskii energy functional. More precisely, they showed that, if $E_N$ denotes the bottom of the spectrum of the operator (\ref{eq:trap-ham}), then
\[ \lim_{N\to \infty} \frac{E_N}{N} = \min_{\ph \in L^2 (\bR^3): \| \ph \|= 1} \eps_{\text{GP}} (\ph) \]
where we defined the Gross-Pitaevskii energy functional \begin{equation}\label{eq:GP-func} \eps_{\text{GP}} (\ph)= \int dx \left( |\nabla \ph (x)|^2 + V_{\text{ext}} (x) |\ph (x)|^2 + 4 \pi a_0 |\ph (x)|^4 \right) \end{equation} for arbitrary $\ph \in H^1 (\bR^3)$. In particular, in first order, the ground state energy depends on the interaction potential $V$ only through its scattering length $a_0$ (the precise profile of the potential is irrelevant). In \cite{LS}, Lieb and Seiringer also showed that the ground state of (\ref{eq:trap-ham}) exhibits complete Bose-Einstein condensation, in the sense that, if $\gamma^{(1)}_N$ denotes the one-particle reduced density matrix associated with the ground state vector $\psi_N$ of (\ref{eq:trap-ham}), then
\begin{equation}\label{eq:cond} \gamma^{(1)}_N \to |\phi_{\text{GP}} \rangle \langle \phi_{\text{GP}}| \, ,  \qquad \text{as } N \to \infty \, , \end{equation}
where $\phi_{\text{GP}} \in L^2(\bR^3)$ denotes the minimizer of the Gross-Pitaevskii energy functional (\ref{eq:GP-func}). Physically, (\ref{eq:cond}) tells us that, in the ground state of (\ref{eq:trap-ham}), almost all particles (all particles up to a fraction vanishing as $N \to \infty$) are in the one-particle state described by  $\phi_{\text{GP}}$.

The analysis of \cite{LSY,LS} implies that if we prepare a trapped Bose gas at sufficiently small temperatures, the system will condensate into the minimizer of the Gross-Pitaevskii energy functional. It seems then natural to ask what happens if we perturb the gas, for example by removing the external traps (typically consisting of strong magnetic fields). The system will immediately react to the perturbation and it will begin to evolve. The evolution will be generated by the translation invariant Hamiltonian
\begin{equation}\label{eq:GP-ham} H_N = \sum_{j=1}^N -\Delta_{x_j} + \sum_{i<j}^N N^2 V (N (x_i - x_j)) \,.\end{equation}
While the results of \cite{LSY,LS} show the validity of the Gross-Pitaevskii theory for predicting the ground state properties of trapped Bose gases, the next theorem shows that it also correctly describes the evolution of initially trapped Bose-Einstein condensates.

\begin{theorem}\label{thm:GP}
Suppose $V \geq 0$, $V(-x) = V(x)$, $|V(x)| \leq (1+|x|^2)^{-\sigma}$, for some $\sigma > 5/2$ and define $H_N$ as in (\ref{eq:GP-ham}). Consider a sequence of $N$-particle wave functions $\psi_N \in L^2 (\bR^{3N})$ such that
\begin{itemize}
\item $\psi_N$ has finite energy per particles; there exists a constant $C >0$ with $\langle \psi_N , H_N \psi_N \rangle \leq C N$.
\item $\psi_N$ exhibits complete condensation; the one-particle reduced density matrix $\gamma^{(1)}_N$ associated with $\psi_N$ is so that $\gamma^{(1)}_N \to |\ph \rangle\langle \ph|$ for some $\ph \in H^1 (\bR^3)$.
\end{itemize}
Let $\psi_{N,t} = e^{-i H_N t } \psi_N$. Then, for every fixed $k \in \bN$ and $t\in \bR$, we have
\[ \tr \left| \gamma^{(k)}_{N,t} - |\ph_t \rangle \langle \ph_t |^{\otimes k} \right| \to 0 \]
as $N \to \infty$, where $\ph_t$ is the solution of the time-dependent nonlinear Gross-Pitaevskii equation
\begin{equation}\label{eq:GP} i\partial_t \ph_t = -\Delta \ph_t + 8 \pi a_0 |\ph_t|^2 \ph_t \end{equation}
with initial data $\ph_{t=0} = \ph$. Here $a_0$ denotes the scattering length of the (unscaled) interaction potential $V$.
\end{theorem}

This theorem, which was proven in \cite{ESY1,ESY2,ESY3,ESY4}, states that complete condensation is preserved by the time-evolution and that the condensate wave function evolves according to the Gross-Pitaevskii equation (\ref{eq:GP}). Note that a different derivation of the Gross-Pitaevskii equation was proposed by Pickl in \cite{P2}.

The Hamilton operator (\ref{eq:GP-ham}) can be written as
\[ H_N = \sum_{j=1}^N -\Delta_{x_j} + \frac{1}{N} \sum_{i<j}^N N^3 V (N (x_i - x_j)) \]
and one may try to interpret $H_N$ as a mean field Hamiltonian, with interaction potential $v_N (x) = N^3 V (N x)$. This observation suggests that, in order to show Theorem \ref{thm:GP}, one can use again the strategy outlined at the end of Section \ref{sec:intro}. It turns out, however, that one should be very careful with this analogy, and that, although the general strategy based on the study of the BBGKY hierarchy still applies, several changes are needed and here a much deeper analysis of the $N$-particle dynamics is required.

Let us briefly explain the new challenges appearing in the proof of Theorem \ref{thm:GP}, as compared with the mean field regime discussed in Section \ref{sec:intro}. Since, formally, $v_N (x) \to b_0 \delta (x)$, with $b_0 = \int dx \, V(x)$, the naive analogy with the mean field situation suggests that the dynamics generated by the Hamiltonian (\ref{eq:GP-ham}) can be approximated, for large $N$, by the Hartree equation
\[ i\partial_t \ph_t = -\Delta \ph_t + \left( b_0 \delta * |\ph_t|^2 \right) \ph_t = -\Delta \ph_t + b_0 |\ph_t|^2 \ph_t \,.\]
This equation has the same form as the Gross-Pitaevskii equation (\ref{eq:GP}), but the wrong coupling constant in front of the non-linearity. {F}rom the physical point of view, it is not surprising that the mean field analogy fails. The mean field regime is characterized by a large number of very weak collisions among the particles. The dynamics generated by (\ref{eq:GP-ham}), on the other hand, is characterized by rare and, at the same time, very strong collisions (particles only interact when they are at distances of order $1/N$ from each others). As a consequence, it turns out that the correlations among the particles, which were negligible in the mean field regime, play here an important role and, in particular, are crucial to understand the emergence of the scattering length $a_0$ in the Gross-Pitaevskii equation (\ref{eq:GP}).

Because of the singularity of the interaction potential, the solution of the Schr\"odinger equation $\psi_{N,t} = e^{-iH_N t} \psi_N$ develops a short-scale correlation structure which lives on the same length scale $1/N$ characterizing the potential. The singular correlation structure can be approximately described by the zero-energy scattering equation $f (x)$ defined in (\ref{eq:0-en}) (more precisely, the singular structure is described by the solution $f_N (x) = f (Nx)$ of the zero-energy scattering equation with rescaled potential $N^2 V (Nx)$). So, the effective, average potential which is experienced say, by particle $j$ due to the interaction with the other $(N-1)$ particles can be approximated here by
\[ \begin{split} \sum_{i \not = j} N^2 V (N(x_i - x_j)) \simeq \; &\sum_{i \not = j} \int dy \, N^2 V (N(y-x_j)) f (N (y - x_j)) |\ph_t (y)|^2 \\ \simeq \; & \int dy V(y) f(y) |\ph_t (x_j + y)|^2 \simeq   8 \pi a_0 |\ph_t (x_j)|^2 \end{split}\] where we used the characterization (\ref{eq:scatt}) of the scattering length $a_0$. Hence, taking into account the correlations, we obtain the Gross-Pitaevskii equation (\ref{eq:GP}) with the correct coupling constant in front of the nonlinearity. This heuristic argument shows that one of the main new challenges consists in proving that the $N$-body dynamics really develops a singular correlation structure which can be described, in good approximation, by the solution of the zero-energy scattering equation (\ref{eq:0-en}). Another major challenge, compared with the results obtained in the mean field regime, is the proof of the uniqueness of the infinite hierarchy. The main problem here is that the interaction potential given in the limiting hierarchy by a delta-function cannot be controlled, in three dimensions, by the kinetic energy. As a consequence, uniqueness is proven in \cite{ESY1} by expanding the solution in a complicated diagrammatic expansion in terms of Feynman graphs; to control the many contributions in this expansion, it is very important to use the smoothing effects of the free evolution, which effectively regularize the singular interaction potential. A new and simpler proof of the uniqueness of the infinite hierarchy was obtained by Klainermann and Machedon in \cite{KM} (later, this approach was extended by Chen and Pavlovi{\'c} in \cite{CP1}). These works show the uniqueness of the infinite hierarchy in a class of densities satisfying certain space-time bounds. Unfortunately, so far it has not been possible to show that the limit points of the sequences $\{ \gamma^{(k)}_{N,t} \}_{k=1}^N$ satisfy these bounds; as a consequence, so far it was not possible to apply the results of \cite{KM,CP1} to prove Theorem \ref{thm:GP} (in one and two dimensions, on the other hand, the results of \cite{KM,CP1} can be applied to show the analogous of Theorem \ref{thm:GP}; see \cite{KSS}). For more details of the proof of Theorem \ref{thm:GP}, we refer to \cite{ESY1,ESY2,ESY3,ESY4}.

\section{Mean Field Evolution of Coherent States}
\label{sec:coh}
\setcounter{equation}{0}

The main drawback of the methods outlined in Sections \ref{sec:intro} and \ref{sec:BEC} is the lack of precise bounds on the difference between the many body dynamics and the effective Hartree evolution. With an expansion of the solution of the BBGKY hierarchy (\ref{eq:BBGKY}), it is possible to show (at least for bounded interaction potentials) that there exists constants $C,T_0 >0$ such that
\begin{equation}\label{eq:stime} \tr \left| \gamma^{(1)}_{N,t} - |\ph_t \rangle \langle \ph_t| \right| \leq \frac{C}{N} \end{equation}
for all $t \in \bR$ with $|t| \leq T_0$. Similar bounds can also be obtained for the reduced $k$-particle densities, for fixed $k \geq 2$. Unfortunately, (\ref{eq:stime}) is only valid for short times; for $t > T_0$, one can still iterate (\ref{eq:stime}), but one only obtains much weaker estimates of the form
\[ \tr \, \left| \gamma^{(1)}_{N,t} - |\ph_t \rangle \langle \ph_t| \right| \leq C N^{-\frac{1}{2^t}} \, . \]

It turns out that one can derive stronger (optimal) bounds on the rate of the convergence of the many body evolution towards the mean field Hartree dynamics using techniques originating in quantum field theory. These techniques, which were first introduced by Hepp in \cite{Hepp} for the analysis of the classical limit of quantum mechanics, and later extended by Ginibre-Velo in \cite{GV}, are based on a Fock-space representation of the bosonic systems and on the study of the dynamics of so-called coherent states. A different approach, which also leads to explicit bounds on the rate of convergence was proposed by Pickl in \cite{P1}, and then applied by Knowles and Pickl to the derivation of the Hartree equation with singular potentials in \cite{KP}.

The bosonic Fock-space over $L^2 (\bR^3)$ is defined as the direct sum
\[ \cF = \bC \oplus \bigoplus_{n \geq 1} L^2_s (\bR^{3n} , dx_1, \dots dx_n) \]
where $L^2_s (\bR^{3n})$ denotes the subspace of $L^2 (\bR^{3n})$ consisting of functions symmetric with respect to permutation of the $n$ particles. Vectors in the Fock-space are sequences $\Psi = \{ \psi^{(n)} \}_{n \geq 0}$, where $\psi^{(n)} \in L^2_s (\bR^{3n})$ is an $n$-particle bosonic wave function. The idea behind the introduction of the Fock-space is that we want to study states where the number of particles is not fixed. Clearly, $\cF$  has the structure of a Hilbert space with the inner product
\[ \langle \Psi , \Phi \rangle = \overline{\psi^{(0)}} \phi^{(0)} + \sum_{n \geq 1} \langle \psi^{(n)} , \phi^{(n)} \rangle \, . \]
The vector $\Omega = \{ 1, 0, 0, \dots \} \in \cF$ is called the vacuum and describes a system with no particles. An important operator on $\cF$ is the number of particle operator $\cN$, which is defined by
\[ \cN \{ \psi^{(n)} \}_{n\geq 0} = \{ n \psi^{(n)} \}_{n\geq 0} \,.\]
The vacuum $\Omega$ is an eigenvector of $\cN$ with eigenvalue zero. More generally, vectors of the form $\{ 0, \dots , 0, \psi^{(m)}, 0, \dots \}$, having a fixed number of particles, are eigenvectors of $\cN$ (with eigenvalue $m$). On $\cF$, we define the Hamilton operator $\cH_N$ by $\cH_N \{ \psi^{(n)} \}_{n \geq 1} = \{ \cH_N^{(n)} \psi^{(n)} \}_{n \geq 1}$, with
\[ \cH_N^{(n)} = \sum_{j=1}^n \left(-\Delta_{x_j} + V_{\text{ext}} (x_j) \right) + \frac{1}{N} \sum_{i<j}^n V (x_i - x_j) \,.\]
By definition, $\cH_N$ leaves each $n$-particle sector $\cF_n$ (defined as the eigenspace of $\cN$ associated with the eigenvalue $n$) invariant. Moreover, on the $N$-particle sector, $\cH_N$ agrees with the mean field Hamiltonian $H_N$ defined in (\ref{eq:mf-ham}). In particular, if we consider the Fock-space evolution of an initial vector with exactly $N$ particles, we find
\[ e^{-i\cH_N t} \{ 0, \dots , 0, \psi_N, 0, \dots \} = \{ 0, \dots, 0, e^{-i H_N t} \psi_N , 0, \dots \} \]
exactly as in Section \ref{sec:intro}. The advantage of working in the Fock space is that we have more freedom in the choice of the initial data. We will use this freedom by considering a class of initial data, known as coherent states, with non-fixed number of particles.

It is very useful to introduce, on the Fock space $\cF$, creation and annihilation operators. For $f \in L^2 (\bR^3)$, we define the creation operator $a^* (f)$ and the annihilation operator $a(f)$ by setting
\[ \begin{split} \left( a^* (f) \psi \right)^{(n)}
(x_1, \dots ,x_n) &= \frac{1}{\sqrt{n}} \sum_{j=1}^n f (x_j)
\psi^{(n-1)} (x_1 , \dots, x_{j-1}, x_{j+1},\dots, x_n) \\ \left( a (f) \psi
\right)^{(n)} (x_1, \dots ,x_n) &= \sqrt{n+1} \int \rd x \;
\overline{f (x)} \psi^{(n+1)} (x, x_1 , \dots , x_n) \, . \end{split} \]
The operators $a^* (f)$ and $a(f)$ are densily defined and closed. It is easy to check that, as the notation suggests, $a^* (f)$ is the adjoint of $a(f)$. Creation and annihilation operators satisfy the canonical commutation relations
\[ \left[ a(f) , a^* (g) \right] = \langle f,g \rangle \qquad \text{and } \quad \left[ a (f) , a (g) \right] = \left[ a^* (f) , a^* (g) \right] = 0 \]
for any $f,g \in L^2 (\bR^3)$ (here $\langle f,g\rangle$ denotes the $L^2$-inner product). Physically, the operator $a^* (f)$ creates a particle with wave function $f$, while $a (f)$ annihilates it. As a consequence, a state with $N$ particles all in the one-particle state $\ph$, can be written as
\[ \{ 0, \dots, 0, \ph^{\otimes N} , 0, \dots \} = \frac{(a^* (\ph))^N}{\sqrt{N!}} \Omega \,. \]
It is also useful to introduce operator-valued distributions $a_x, a^*_x$ defined so that
\[ a(f) = \int dx \overline{f (x)} a_x, \qquad \text{and } \quad a^* (f) = \int dx \, f(x) \, a^*_x \, .\]
With this notation, $a_x^* a_x$ gives the density of particles close to $x\in \bR^3$. The number of particles operator can formally be written as
\[ \cN = \int dx \, a_x^* a_x \, .\]
Similarly, the Hamilton operator $\cH_N$ can be formally expressed in terms of operator-valued distributions as
\[ \cH_N = \int dx \, \left( \nabla_x a^*_x \, \nabla_x a_x + V_{\text{ext}} (x) a_x^* a_x \right) +  \frac{1}{2N} \int dx dy \, V (x-y) a_x^* a_y^* a_y a_x \,. \]
The fact that every term in the Hamiltonian contains the same number of creation and annihilation operators means that $\cH_N$ commutes with the number of particles or, equivalently, that the number of particles is preserved by the time-evolution enerated by $\cH_N$.

For later use, we observe that the creation and annihilation operators are not bounded; however, they can be bounded by the square root of the number of particles operator, in the sense that
\begin{equation}\label{eq:bd-aa*}
\begin{split}\| a(f) \psi \| &\leq \| f \| \, \| \cN^{1/2} \psi \| \qquad \text{and } \qquad  \| a^* (f) \psi \| \leq \| f \| \, \| (\cN +1)^{1/2} \psi \|
\end{split}\end{equation}
for every $\psi \in \cF$, $f \in L^2 (\bR^3)$ (here $\| f \|$ indicates the $L^2$-norm of $f$).

As mentioned above, we are going to study the evolution of initial coherent states.
For arbitrary $\ph \in L^2 (\bR^3)$, we define the Weyl operator
\[ W (\ph) = e^{a^* (\ph) - a(\ph)} \,.\]
The coherent state with wave function $\ph$ is then defined as $W(\ph) \Omega$. The Weyl operator $W(\ph)$ is a unitary operator; therefore the coherent state $W(\ph) \Omega$ always has norm one. Moreover, since
\[ W (\ph) \Omega = e^{-\| \ph \|^2/2} \sum_{j=0} \frac{a^* (\ph)^j}{j!} \Omega = e^{-\| \ph \|^2/2} \{ 1, \ph, \frac{\ph^{\otimes 2}}{\sqrt{2!}}, \dots, \frac{\ph^{\otimes j}}{\sqrt{j!}}, \dots \} \, , \]
the coherent state $W(\ph) \Omega$ does not have a fixed number of particles. One can nevertheless compute the expectation of the number of particles in the state $W (\ph) \Omega$; it is given by
\[ \langle W (\ph) \Omega, \cN W (\ph) \Omega \rangle = \| \ph \|^2 \, . \]
More precisely, it turns out that the number of particle in the coherent state $W(\ph) \Omega$ is a Poisson random variable with expectation and variance $\| \ph \|^2$. The main reason why coherent states have nice algebraic properties (which will be used later on in the analysis of their evolution) is the fact that they are eigenvectors of all annihilation operators. Indeed
\[ a (f) W (\ph) \Omega = (f, \ph) W(\ph) \Omega \]
for every $f,\ph \in L^2 (\bR^3)$. This is a simple consequence of the fact that Weyl operators generate shifts of creation and annihilation operators, in the sense that
 \begin{equation}\label{eq:shift} W^* (\ph) a (f) W(\ph) = a (f) + (f, \ph), \qquad \text{and } \quad W^* (\ph) a^* (f) W(\ph) = a^* (f) + (\ph , f) \, . \end{equation}

Next, we study the evolution of an initial coherent state with respect to the dynamics generated by the Hamilton operator $\cH_N$. To reproduce the mean field regime analyzed in Section \ref{sec:intro}, we choose the initial coherent state to have expected number of particles equal to $N$ (the number of particles cannot be fixed, but at least we should fix its average to be given by $N$; otherwise the resulting evolution will not have anything to do with the mean field Hartree dynamics).

\begin{theorem}\label{thm:coh}
Suppose that the interaction potential $V$ is such that, as an operator inequality $V^2 \leq (1-\Delta)$. For $\ph \in H^1 (\bR^3)$, consider the initial coherent state
\[ W (\sqrt{N} \ph) \Omega = e^{-N/2} \left\{ 1, \sqrt{N} \ph , \dots , \frac{N^{j/2} \ph^{\otimes j}}{\sqrt{j!}}, \dots \right\}\,. \]
Let $\Psi_{N,t} = e^{-i\cH_N t} W(\sqrt{N} \ph) \Omega$, and let $\Gamma^{(1)}_{N,t}$ denote the one-particle reduced density associated with $\Psi_{N,t}$. Then there exist constants $C,D \geq 0$ such that
\begin{equation}\label{eq:bd-coh} \tr\, \left| \Gamma^{(1)}_{N,t} - |\ph_t \rangle \langle \ph_t| \right| \leq C \frac{e^{D |t|}}{N} \end{equation}
for all $t \in \bR$. Here $\ph_t$ denotes the solution of the nonlinear one-particle Hartree equation (\ref{eq:hartree}) with initial data $\ph_{t=0} = \ph$.
\end{theorem}

Observe that the operator inequality $V^2 \leq C (1-\Delta)$, which means that
\[ \int dx \, V(x) |\ph (x)|^2 \leq C \| \ph \|_{H^1} \]
for all $\ph \in H^1 (\bR^3)$, is satisfied for potentials with Coulomb singularities $V(x) \simeq \pm 1 / |x|$.

{F}rom the convergence towards the Hartree dynamics for the evolution of initial coherent states, one can deduce a similar result for the evolution of initially factorized states with a fixed number of particles. To this end, one can use the fact that
\[ \{ 0, \dots , 0, \ph^{\otimes N} , 0 , \dots \} = \frac{a^* (\ph)^N}{\sqrt{N!}} \Omega = d_N P_N W(\sqrt{N} \ph) \Omega \]
where $P_N$ is the orthogonal projection onto the $N$-particle sector of $\cF$, and $d_N = e^{N/2} N^{-N/2} \sqrt{N!} \simeq N^{1/4}$. Alternatively, one can write
\[ \{ 0, \dots , 0, \ph^{\otimes N} , 0 , \dots \}= d_N \int_0^{2\pi} d\theta \, e^{-i N\theta} W(\sqrt{N} e^{i\theta} \ph) \Omega \] to express the factorized state as a linear combination of coherent states.

\begin{corollary}\label{cor}
Let the potential $V$ be so that $V^2 \leq (1-\Delta)$ and define the Hamiltonian $H_N$ as in (\ref{eq:mf-ham}). Let $\psi_N = \ph^{\otimes N}$ and $\psi_{N,t} = e^{-i H_N t} \psi_N$. Then there exist $C,D >0$ such that
\[ \tr\, \left| \gamma^{(1)}_{N,t} - |\ph_t \rangle \langle \ph_t| \right| \leq C \frac{e^{D|t|}}{N}  \]
for all $t \in \bR$. Here $\ph_t$ is the solution of the Hartree equation (\ref{eq:hartree}), with initial data $\ph_{t=0} = \ph$.
\end{corollary}

The details of how Corollary \ref{cor} follow from Theorem \ref{thm:coh} can be found in \cite{RS,CLS}. In the following let us briefly present the main ideas behind the proof of Theorem \ref{thm:coh}. The main idea is to use the fact that the evolution of a coherent state remains approximately coherent. As we will see, it is possible to extract the coherent part of the evolved state, and then to focus on the evolution of the fluctuations, which, thank to the algebraic properties of the coherent states, can be expressed in simple and compact form. Let us remark that, in \cite{GMM1,GMM2}, Grillakis, Machedon and Margetis proposed to approximate the evolution of the coherent state not just by a coherent state but by a larger manifold of so-called Bogoliubov states; this allows them to obtain more precise approximation of the many-body evolution (this approach has been extended to systems with three-body interactions by Chen in \cite{C}).

The first observation is that the one-particle density matrix $\Gamma^{(1)}_{N,t}$ associated with the Fock-space state $\Psi_{N,t}$ has the integral kernel
\[ \Gamma^{(1)}_{N,t} (x;y) = \frac{1}{\langle \Psi_{N,t}, \cN \Psi_{N,t} \rangle} \, \langle \Psi_{N,t}, a^*_x a_y \Psi_{N,t} \rangle = \frac{1}{N}  \langle \Psi_{N,t}, a^*_x a_y \Psi_{N,t} \rangle \,.\]
Expanding $a_x^*$ and $a_y$ around their mean field values $\sqrt{N} \ph_t (x)$, $\sqrt{N} \ph_t (y)$, we obtain
\[ \begin{split} \Gamma^{(1)}_{N,t} (x;y) &- \overline{\ph}_t (x) \ph_t (y) \\ = & \; \frac{1}{N}  \left\langle \Omega, W^* (\sqrt{N} \ph) e^{i\cH_N t} \left(a^*_x - \sqrt{N} \overline{\ph}_t (x) \right) \left( a_y -\sqrt{N} \ph_t (y) \right) e^{-i\cH_N t} W (\sqrt{N} \ph) \Omega \right\rangle \\ & + \frac{\overline{\ph}_t (x)}{\sqrt{N}} \left\langle \Omega, W^* (\sqrt{N} \ph) e^{i\cH_N t} \left( a_y -\sqrt{N} \ph_t (y) \right) e^{-i\cH_N t} W (\sqrt{N} \ph) \Omega \right\rangle \\ & + \frac{\ph_t (y)}{\sqrt{N}} \left\langle \Omega, W^* (\sqrt{N} \ph) e^{i\cH_N t} \left( a^*_x -\sqrt{N} \overline{\ph}_t (x) \right) e^{-i\cH_N t} W (\sqrt{N} \ph) \Omega \right\rangle \,. \end{split} \]
Using (\ref{eq:shift}), we rewrite the last equation as
\[\begin{split} \Gamma^{(1)}_{N,t} (x;y) &- \overline{\ph}_t (x) \ph_t (y) \\ = & \; \frac{1}{N}  \left\langle \Omega, W^* (\sqrt{N} \ph) e^{i\cH_N t} W (\sqrt{N} \ph_t) a^*_x a_y W^* (\sqrt{N} \ph_t) e^{-i\cH_N t} W (\sqrt{N} \ph) \Omega \right\rangle \\ & + \frac{\overline{\ph}_t (x)}{\sqrt{N}} \left\langle \Omega, W^* (\sqrt{N} \ph) e^{i\cH_N t} W (\sqrt{N} \ph_t) a_y W^* (\sqrt{N} \ph_t) e^{-i\cH_N t} W (\sqrt{N} \ph) \Omega \right\rangle \\ & + \frac{\ph_t (y)}{\sqrt{N}} \left\langle \Omega, W^* (\sqrt{N} \ph) e^{i\cH_N t} W(\sqrt{N} \ph_t) a^*_x  W^* (\sqrt{N} \ph_t) e^{-i\cH_N t} W (\sqrt{N} \ph) \Omega \right\rangle \,.\end{split} \]
Introducing the two-parameter group of unitary transformations \begin{equation}
\label{eq:cU} \cU (t;s) = W^* (\sqrt{N} \ph_t) e^{-i \cH_N (t-s)} W (\sqrt{N} \ph_s),\end{equation}
we find
\begin{equation}\label{eq:Gamma1} \begin{split} \Gamma^{(1)}_{N,t} (x;y) - \overline{\ph}_t (x) \ph_t (y) = & \; \frac{1}{N}  \langle \Omega, \cU^* (t;0) \, a^*_x a_y  \, \cU (t;0) \Omega \rangle  + \frac{\overline{\ph}_t (x)}{\sqrt{N}} \langle \Omega, \cU^* (t;0) a_y \, \cU (t;0) \Omega \rangle  \\ &+ \frac{\ph_t (y)}{\sqrt{N}} \langle \Omega, \cU^* (t;0) a^*_x \, \cU (t;0) \Omega \rangle \, .\end{split} \end{equation}
Let us first consider the first term on the r.h.s. of the last equation; recalling the bounds (\ref{eq:bd-aa*}), we conclude that the contribution of this term to the l.h.s. of (\ref{eq:bd-coh}) can be controlled by the r.h.s. of (\ref{eq:bd-coh}) if we can control the growth of the expectation of the number of particles operator $\cN$ with respect to the fluctuation dynamics $\cU (t;0)$, i.e. if we can prove that
\begin{equation}\label{eq:growN} \langle \Omega, \cU^* (t;0) \, \cN \cU (t;0) \Omega \rangle \leq C e^{D |t|} \, .\end{equation}
It is worth noticing that the fluctuation dynamics $\cU (t;s)$ satisfies the Schr\"odinger equation
\[ i\partial_t \cU (t;s) = \cL(t) \cU (t;s), \qquad \text{with } \cU (s;s) = 1 \]
with the generator
\begin{equation}\label{eq:L} \begin{split} \cL (t) = \; & \int dx \, \left(\nabla_x a^*_x \nabla_x a_x + V_{\text{ext}} (x) \, a_x^* a_x + (V*|\ph_t|^2) (x) a_x^* a_x \right) + \int dx dy \, V(x-y) \, \ph_t (x) \overline{\ph}_t (y) a_x^* a_y \\ & + \frac{1}{2} \int dx dy \, V(x-y) \, \left( \ph_t (x) \ph_t (y) a^*_x a^*_y + \overline{\ph}_t (x) \overline{\ph}_t (y) a_x a_y \right) \\ &+\frac{1}{2 \sqrt{N}} \int dx dy V (x-y) \, a_x^* \left( a_y^* \ph_t (y) + a_y \overline{\ph}_t (y) \right) a_x \\ &+\frac{1}{2N} \int dx dy \, V( x-y) \, a_x^* a_y^* a_y a_x \,.\end{split} \end{equation}
In contrast with the Hamiltonian $\cH_N$, the generator $\cL (t)$ contains terms (the terms on the second and third line) which do not commute with the number of particles operator $\cN$ (because the number of creation operators does not match the number of annihilation operators). Hence, not surprisingly, the expectation of $\cN$ is not preserved by the fluctuation dynamics $\cU (t;0)$. Nevertheless, it turns out that, if the condition $V^2 \leq C (1-\Delta)$ is satisfied, it is still possible to control the growth of the expectation of $\cN$, and to prove the bound (\ref{eq:growN}). For the details, we refer to \cite{RS}.

The analysis of the second and the third term on the r.h.s. of (\ref{eq:Gamma1}) requires slightly more works. Formally these terms seem to be of the order $N^{-1/2}$. To show that they really are of the order $N^{-1}$, and prove the bound (\ref{eq:bd-coh}), one needs to compare the fluctuation dynamics $\cU (t;0)$ with a limiting dynamics $\cU_{\infty} (t;0)$, satisfying the Schr\"odinger equation
\[ i\partial_t \cU_{\infty} (t;s) = \cL_{\infty} (t) \cU_{\infty} (t;s) \]
with the generator
\[ \begin{split} \cL_{\infty} (t) = \; & \int dx \, \left(\nabla_x a^*_x \nabla_x a_x + V_{\text{ext}} (x) \, a_x^* a_x + (V*|\ph_t|^2) (x) a_x^* a_x \right) + \int dx dy \, V(x-y) \, \ph_t (x) \overline{\ph}_t (y) a_x^* a_y \\ & + \frac{1}{2} \int dx dy \, V(x-y) \, \left( \ph_t (x) \ph_t (y) a^*_x a^*_y + \overline{\ph}_t (x) \overline{\ph}_t (y) a_x a_y \right) \end{split} \] obtained by formally letting $N \to \infty$ in (\ref{eq:L}). The point is that if we replace $\cU (t;0)$ by $\cU_{\infty} (t;0)$ in the second and in the third term on the r.h.s. of (\ref{eq:Gamma1}), these terms vanish. In fact
\[ \langle \Omega, \cU_{\infty} (t;0)^* a_x \, \cU_{\infty} (t;0) \Omega \rangle =  \langle \Omega, \cU_{\infty} (t;0)^* a^*_x \, \cU_{\infty} (t;0) \Omega \rangle = 0 \] because, although $\cU_{\infty} (t;0)$ does not preserves the number of particles, it does preserve the parity (in the sense that it commutes with the operator $(-1)^{\cN}$). This observation implies that, in the second and third terms on the r.h.s. of (\ref{eq:Gamma1}), the unitary evolution $\cU (t;0)$ can be replaced by the difference $\cU (t;0) - \cU_{\infty} (t;0)$; this produces the additional factor $N^{-1/2}$ (because the difference between the two generators, $\cL (t)$ and $\cL_{\infty} (t)$, is of this order), and explains why also the second and the third term on the r.h.s. of (\ref{eq:Gamma1}) are of the order $N^{-1}$. Again, we refer to \cite{RS} for further details.

\section{Gravitational Collapse of Boson Stars}
\label{sec:grav}
\setcounter{equation}{0}

As an application of the techniques discussed in Section \ref{sec:coh}, the last part of these notes is devoted to the study of the dynamics of so-called boson stars. These are systems of bosons with relativistic dispersion law interacting through classical Newtonian gravity (such particles are usually called semi-relativistic, or pseudo-relativistic bosons). The Hamilton operator has the form
\[ H_N^{\text{grav}} = \sum_{j=1}^N \sqrt{1-\Delta_{x_j}} - G \sum_{i<j}^N \frac{1}{|x_i - x_j|} \,.\]
We are interested here in the mean field regime, characterized by $N \gg 1$ and $G \ll 1$, so that $\kappa := NG$ remains fixed, of order one. Since, in the units we use, the gravitational constant $G$ is approximately given by $G \simeq 10^{-45}$, this means that we are interested in systems with approximately $N \simeq 10^{45}$ particles. To analyze this regime, we consider the time-evolution generated by
\begin{equation}\label{eq:grav-ham} H_N = \sum_{j=1}^N \sqrt{1-\Delta_{x_j}} - \frac{\kappa}{N} \sum_{i<j}^N \frac{1}{|x_i -x_j|} \end{equation}
in the limit $N \to \infty$. Inspired by the results discussed in the previous sections, we expect that, in this limit, the evolution generated by $H_N$ on factorized initial data can be approximated by the mean field semi-relativistic Hartree equation
\begin{equation}\label{eq:semi} i\partial_t \ph_t = \sqrt{1-\Delta} \, \ph_t - \kappa \left( \frac{1}{|.|} * |\ph_t|^2 \right) \ph_t \,. \end{equation}

It turns out that the system under consideration is critical. This follows from the observation that, for large momenta, the kinetic energy $\sqrt{1-\Delta} \simeq |\nabla|$ scales like the potential energy, as an inverse length. This implies that, for arbitrary $N \in \bN$, there exists a critical coupling $\kappa_{\text{cr}} (N) >0$ such that
\[ \inf_{\psi_N \in L^2 (\bR^{3N}), \| \psi_N \| = 1}  \langle \psi_{N} , H_N \psi_{N} \rangle \geq 0 \]
for $\kappa \leq \kappa_{\text{cr}} (N)$, and
\[ \inf_{ \psi_N \in L_s^2 (\bR^{3N}), \| \psi_N \| = 1} \langle \psi_N , H_N \psi_N \rangle = -\infty \]
for $\kappa > \kappa_{\text{cr}}$. In other words, for small $\kappa >0$, the kinetic energy controls the potential energy, and the total energy is bounded below. For $\kappa > \kappa_{\text{cr}}$, on the other hand, the potential energy dominates and leads the total energy to arbitrary negative values. Criticality can also be observed on the level of the semi-relativistic Hartree equation (\ref{eq:semi}). As proven in \cite{L}, (\ref{eq:semi}) is locally well-posed in the energy space $H^{1/2} (\bR^3)$ for arbitrary coupling constants. Its global behavior, however, depends on the value of $\kappa$. There exists namely a critical coupling $\kappa_{\text{cr}} >0$ with the following properties. For $\kappa < \kappa_{\text{cr}}$, (\ref{eq:semi}) is globally well-posed in $H^{1/2} (\bR^3)$, in the sense that all local solutions extend to all times. For $\kappa > \kappa_{\text{cr}}$, on the other hand, (\ref{eq:semi}) has local solutions $\ph_t \in C ([0,T) , H^{1/2} (\bR^3))$ exhibiting blow up in finite time, in the sense that
\[ \| \ph_t \|_{H^{1/2}} \to \infty \] as $t \to T^-$. In \cite{FL}, Fr\"ohlich and Lenzmann proved that arbitrary spherically symmetric initial data with negative energy lead, if $\kappa > \kappa_{\text{cr}}$ to blow up in finite time (the spherical symmetry is believed to be just a technical condition). In the physics literature, the blow up solutions of the semi-relativistic Hartree equation (\ref{eq:semi}) have been used to describe the phenomenon of stellar or gravitational collapse, first predicted by Chandrasekhar.

Mathematically, it seems important to understand whether the relation between the many body evolution and the semi-relativistic Hartree dynamics (\ref{eq:semi}) can be established rigorously. The first step in this direction was accomplished by Lieb and Yau; in \cite{LY}, they proved that $\kappa_{\text{cr}} (N) \to \kappa_{\text{cr}}$ as $N \to \infty$, and that, for $\kappa < \kappa_{\text{cr}}$, the ground state energy per particle converges, as $N \to \infty$, to the minimum of the Hartree energy functional
\[ \eps_{\text{Hartree}} (\ph) = \int dx \, \left| (1-\Delta)^{1/4} \ph \right|^2 - \frac{\kappa}{2} \int dx dy \frac{|\ph (x)|^2 |\ph (y)|^2}{|x-y|} \]
over all $\ph \in L^2 (\bR^3)$ with $\| \ph \| =1$. This proves that the Hartree theory successfully predicts the ground state properties of the boson star. Does it also describe their time-evolution?

A first positive answer to this question was obtained in \cite{ES}, for the subcritical case $\kappa < \kappa_{\text{cr}}$. To consider the supercritical case $\kappa > \kappa_{\text{cr}}$, we first have to give a precise mathematical definition of the
many body evolution. In fact, for $\kappa > \kappa_{\text{cr}} (N)$, it is not so simple to define the many body evolution because $H_N$ is not bounded below. As a consequence, $H_N$ does not have a unique and natural extension as a self-adjoint operator on the Hilbert space $L^2 (\bR^{3N})$. Therefore, it is not clear how to define the unitary group $e^{-i H_N t}$ and it is not clear how to solve the many body Schr\"odinger equation. A possible way to avoid this problem is to consider weak solutions of the many body Schr\"odinger equation. Here, we follow a different approach. To circumvent the fact that the Hamiltonian is unbounded from below, we introduce an arbitrarily small, $N$-dependent cutoff $\alpha_N$ in the interaction, defining the regularized Hamiltonian
\[ H_N^{\alpha} = \sum_{j=1}^N \sqrt{1-\Delta_{x_j}} - \frac{\kappa}{N} \sum_{i<j} \frac{1}{|x_i -x_j| + \alpha_N} \,. \]
The cutoff $\alpha_N$ is assumed to be strictly positive for all $N \in \bN$, and to vanish in the limit of large $N$. For arbitrary $N \in \bN$, $H_N^{\alpha}$ is now bounded below and generates therefore a well-defined unitary evolution on $L^2 (\bR^{3N})$. On the other hand, since $\alpha_N$ vanishes as $N \to \infty$, we do not expect it to considerably affect the macroscopic properties of the dynamics. {F}rom the physical point of view, the introduction of the $N$-dependent cutoff is justifiable by the fact that, anyway, at very small distances, Newton's gravity is effectively regularized by the presence of other forces.

Now, we are ready to study the relation between the many body evolution generated by $H_N^\alpha$ and the semi-relativistic Hartree dynamics (\ref{eq:semi}). The next theorem, proven in \cite{MS}, shows that, if the nonlinear dynamics is well defined in a time interval $[-T,T]$, then, in this time interval, (\ref{eq:semi}) really approximates the many body evolution in the limit of large $N$.

\begin{theorem}\label{thm:MS1}
Let $\ph \in H^2 (\bR^3)$, $\alpha_N >0$ such that $\alpha_N \to 0$ as $N \to \infty$, $\psi_{N,t} = e^{-i H_N^{\alpha} t} \ph^{\otimes N}$. Let $\ph_t$ be the solution of the semi-relativistic Hartree equation (\ref{eq:semi}), with initial data $\ph_{t=0} = \ph$. Fix $T > 0$, and assume that
\[ \lambda := \sup_{|t| \leq T} \| \ph_t \|_{H^{1/2} (\bR^3)} < \infty \, . \]
Then, for every $k\in \bN$, there exists a constant $C = C (k,T,\lambda) > 0$ such that
\begin{equation} \label{eq:conv-grav} \sup_{|t| \leq T} \, \tr \, \left| \gamma^{(k)}_{N,t} - |\ph_t \rangle \langle \ph_t|^{\otimes k} \right| \leq C \left( \frac{1}{N^{1/2}} + \alpha_N \right) \, . \end{equation}
\end{theorem}

What happens now if the Hartree dynamics exhibits blow-up. The next theorem, which was also proven in \cite{MS}, shows that, if the Hartree dynamics blows up at time $T$, then also the many body evolution collapses if $t \to T$, and, simultaneously, $N \to \infty$ (at a sufficient fast rate). By collapse of the solution $\psi_{N,t}$ of the many body Schr\"odinger equation, we mean the following. The kinetic energy per particle, given by
\[ \langle \psi_{N,t} , (1-\Delta_{x_1})^{1/2} \psi_{N,t} \rangle = \tr \, (1-\Delta)^{1/2} \gamma^{(1)}_{N,t} \, , \]
is finite, uniformly in $N$, for all $t <T$, while it diverges to infinity if $t \to T$ and $N \to \infty$.

\begin{theorem}\label{thm:MS2}
Assume that, in the definition of $H_N^\alpha$, $\alpha_N \geq N^{-\ell}$ for some (arbitrarily large) $\ell >0$. Let $\ph \in H^2 (\bR^3)$, $\psi_{N,t} = e^{-i H_N^{\alpha} t} \ph^{\otimes N}$. Let $\ph_t$ be the solution of the semi-relativistic Hartree equation (\ref{eq:semi}) with initial data $\ph_{t=0} = \ph$. Assume that $\ph_t$ is locally well posed in $H^{1/2} (\bR^3)$ in the interval $[0,T)$ and that it blows up at time $T$, that is $\| \ph_t \|_{H^{1/2}} < \infty$, for all $t\in [0,T)$ and
\[ \lim_{t \to T^-} \| \ph_t \|_{H^{1/2}} = \infty \, .\]
Then, for every $0 \leq t < T$, we have \begin{equation}\label{eq:MS21} \tr\, (1-\Delta)^{1/2} \gamma^{(1)}_{N,t} < \infty \end{equation} uniformly in $N$. Moreover, if $N(t) \to \infty$ as $t \to T^-$ sufficiently fast, we have
\begin{equation}\label{eq:MS22} \lim_{t \to T^-} \tr \, (1-\Delta)^{1/2} \gamma^{(1)}_{N(t),t} = \infty \,.\end{equation}
\end{theorem}

This theorem establishes the (dynamical) gravitational collapse of the boson star directly on the level of the many body evolution, justifying the use of the semi-relativistic Hartree equation (\ref{eq:semi}) for the description of the dynamics.

The proof of Theorems \ref{thm:MS1} and \ref{thm:MS2} relies on the ideas discussed in Section \ref{sec:coh}. In particular, we use a Fock-space representation of the system and we study the evolution of initial coherent states. The main new challenge is that, in order to prove Theorem \ref{thm:MS2} (in particular, in order to show (\ref{eq:MS21})), we establish the convergence of the one-particle reduced density $\Gamma^{(1)}_{N,t}$ associated with the evolution of the initial coherent state towards $|\ph_t \rangle \langle \ph_t|$ in an energy norm (in Section \ref{sec:coh}, convergence was established in the trace norm, which, in the language of density matrices, is the equivalent of an $L^2$-norm). More precisely, we show that, for any $t_0 < T$ (recall that $T$ is the blow-up time of the nonlinear Hartree equation (\ref{eq:semi})),
\begin{equation}\label{eq:en-conv} \sup_{t \in [0, t_0]} \tr \left| (1-\Delta)^{1/4} \left( \Gamma^{(1)}_{N,t} - |\ph_t \rangle\langle \ph_t| \right) (1-\Delta)^{1/4} \right| \leq C N^{-1/2} \end{equation} where the constant $C>0$ depends only on $t_0$ and on $\sup_{t \in [0,t_0]} \| \ph_t \|_{H^{1/2}}$. In Section \ref{sec:coh}, the proof of the convergence of the reduced density matrix $\Gamma^{(1)}_{N,t}$ in the trace norm reduced to the problem of controlling the growth the number of particles operator with respect to the fluctuation dynamics $\cU (t;0)$ defined in (\ref{eq:cU}). Similarly, the proof of (\ref{eq:en-conv}) reduces to the problem of controlling the growth of the expectation of the (relativistic) kinetic energy operator with respect to the fluctuation dynamics.
The fundamental reason why this is possible is that, after factoring out the (super-critical) Hartree dynamics from the (super-critical) many body evolution, the dynamics of the fluctuation is sub-critical. For further details, we refer to \cite{MS}.

\end{document}